# Chebyshev Approximation and Higher Order Derivatives of Lyapunov Functions for Estimating the Domain of Attraction

Dongkun Han and Dimitra Panagou

*Abstract*— Estimating the Domain of Attraction (DA) of non-polynomial systems is a challenging problem. Taylor expansion is widely adopted for transforming a nonlinear analytic function into a polynomial function, but the performance of Taylor expansion is not always satisfactory. This paper provides solvable ways for estimating the DA via Chebyshev approximation. Firstly, for Chebyshev approximation without the remainder, higher order derivatives of Lyapunov functions are used for estimating the DA, and the largest estimate is obtained by solving a generalized eigenvalue problem. Moreover, for Chebyshev approximation with the remainder, an uncertain polynomial system is reformulated, and a condition is proposed for ensuring the convergence to the largest estimate with a selected Lyapunov function. Numerical examples demonstrate that both accuracy and efficiency are improved compared to Taylor approximation.

## I. INTRODUCTION

Stability of a single equilibrium point is of great significance in the analysis of dynamical systems. Fortunately, the stability of systems with respect to a specific initial point can be directly predicted if we know the domain of attraction (DA), i.e., a set of initial points from which all trajectories converge to the equilibrium point. Estimating the DA stems from the transient stability problem of electrical systems, and more applications appear recently in various areas, e.g., biological systems, chemical reactions, intelligent transportations, just to name a few (good surveys can be found in [1], [2]).

To obtain the exact DA, elegant methods have been proposed based on *Zubov-like equation* and *maximal Lyapunov functions*, but the solution of Zubov-like equation and the maximal Lyapunov function are not easy to find [3], [4]. In order to solve this issue, methods based on sublevel sets of Lyapunov functions are proven to be an efficient way, thanks to the recent development in semidefinite programming and sum-of-squares technique [5]. Effective methods have been proposed with different types of Lyapunov functions, including *quadratic Lyapunov functions* [6], *polyhedral Lyapunov functions* [7], *polynomial Lyapunov functions* [8], *pointwise maximum Lyapunov functions* [9], and more recently *rational polynomial Lyapunov functions* [10]. On the other hand, various methods have been provided to decrease the conservativeness of Lyapunov function methods [11]–[14]. In [15], an *invariant set* is formulated by searching a Lyapunov function. In [16], conditions based on higher order derivatives of a Lyapunov-like function are proposed for constructing invariant sets. In [17], [18], non-monotonic Lyapunov functions are studied and an efficient method is proposed based on finding a group of polynomials rather than a standard Lyapunov function. These methods are effective for quadratic or polynomial vector fields.

However, demands for the analysis of non-polynomial systems are increasing in numerous situations, e.g., robotic limbs control, and aircraft navigation in longitude flight [19]. To cope with non-polynomial systems, elegant methods are proposed based on polynomial approximation, such as state space recasting [19], and polynomials inclusion [20]. In [21], a method is proposed by using the truncated Taylor expansion, and the largest estimate of the DA is obtained by using polynomial Lyapunov functions. Based on this idea, the DA of rational non-polynomial systems is computed, also using the truncated Taylor expansion [22].

Inspired by the work in [18], in contrast to our previous work [22], [23] using Taylor approximation, this paper deploys higher order derivatives of Lyapunov functions via Chebyshev approximation. The main contributions are displayed as follows:

- For Chebyshev approximation without the remainder, higher order derivatives of Lyapunov functions are used for estimating the DA, a sufficient condition is proposed via Real Positivestellensatz. Then, by using the technique of Squared Matrix Representation (SMR), the largest estimate is achieved by solving a generalized eigenvalue problem.
- For Chebyshev approximation with the remainder, an uncertain polynomial system is reformulated, and a condition is proposed for ensuring the convergence to the largest estimate with a selected Lyapunov function. Results for the optimal Lyapunov function is also provided by using sum-of-squares programming.

The rest of the paper is organized as follows. Preliminary knowledge on Chebyshev approximation is briefly introduced in Section II. Solvable conditions are proposed for Chebyshev approximation with and without the remainder in Section III. In Sections IV, the advantages of the proposed method are shown by several examples. Finally, conclusions and future work are discussed in Section V.

The authors are with the Department of Aerospace Engineering, University of Michigan, Ann Arbor, MI, USA. E-mail: {dongkunh, dpanagou}@umich.edu.

This work was sponsored by the Automotive Research Center (ARC) in accordance with Cooperative Agreement W56HZV-14-2-0001 U.S. Army TARDEC in Warren, MI, USA, the NASA Grant NNX16AH81A, and an Early Career Faculty Grant from NASA's Space Technology Research Grants Program.

## II. PRELIMINARIES

Notations: $\mathbb{N}, \mathbb{R}$: natural and real number sets; $\mathbb{R}_+$: positive real number set; $0_n$: origin of $\mathbb{R}^n$; $\mathbb{R}_0^n$: $\mathbb{R}^n \setminus \{0_n\}$; $A^T$: transpose of $A$; $A > 0$ ($A \geq 0$): symmetric positive definite (semidefinite) matrix $A$; $A \otimes B$: Kronecker product of matrices $A$ and $B$; $\deg(f)$: degree of polynomial function $f$; $\text{trace}(A)$: trace of matrix $A$; $(*)^T A B$ in a form of SMR: $B^T A B$. Let $\mathcal{P}$ be the set of polynomials and $\mathcal{P}^{n \times m}$ be the set of matrix polynomials with dimension $n \times m$. A polynomial $p(x) \in \mathcal{P}$ is nonnegative if $p(x) \geq 0$ for all $x \in \mathbb{R}^n$. A useful way of establishing $p(x) \geq 0$ consists of checking whether $p(x)$ can be described as a sum of squares of polynomials (SOS), i.e., $p(x) = \sum_{i=1}^{k} p_i(x)^2$ for some $p_1, \ldots, p_k \in \mathcal{P}$. The set of SOS polynomials is denoted by $\mathcal{P}^{\text{SOS}}$. If $p(x) \in \mathcal{P}^{\text{SOS}}$ becomes 0 only for $x = 0_n$, we call $p(x)$ *local SOS* denoted by $\mathcal{P}_0^{\text{SOS}}$.

### A. Model Formulation

In this paper, we consider the following autonomous dynamical system:

$$\dot{x}(t) = f(x(t)) + \sum_{i=1}^{r} g_i(x(t)) \zeta_i(x_{a_i}(t)), \quad (1)$$

where $x \in \mathbb{R}^n$ is the state vector, $x(0) \in \mathbb{R}^n$ is the initial state, $\chi(t; x(0))$ denotes the solution of system (1), with the polynomial functions $f \in \mathcal{P}^n, g_1, \ldots, g_r \in \mathcal{P}$, the non-polynomial functions $\zeta_1, \ldots, \zeta_r : \mathbb{R} \to \mathbb{R}^n$ and the indexes $a_1, \ldots, a_r \in \{1, \ldots, n\}$. In order to use $k$th-order Chebyshev approximation, we assume that $\zeta_i$ is $k$ times differentiable, where $k$ is called the truncation degree. In the sequel, we will omit the arguments $t$ and $x$ of functions whenever possible for the brevity of notations.

*Remark 1:* Without considering the Chebyshev remainder, non-polynomial functions $\zeta_i$ only require to be continuous. However, for Chebyshev approximation with the remainder and the truncation degree $k$, we assume $\zeta_i$ to be $k$ times differentiable (For details, please see the expression of remainder in Section III.B). □

*Remark 2:* In this work, we are concerned with the DA of a single stable equilibrium. Without loss of generality, we set the equilibrium as the origin. □

The DA of the origin is expressed as:

$$\mathcal{D} = \left\{ x(0) \in \mathbb{R}^n : \lim_{t \to +\infty} \chi(t; x(0)) = 0_n \right\},$$

where $\chi$ is the solution of system (1).

### B. Chebyshev Approximation

An essential observation in [24] is that the convergence of Taylor series is not always ensured. The following example is provided for illustration.

*Example 1:* Let

$$\zeta(x) = \begin{cases} e^{-\frac{1}{x^2}}, & \text{for } x \neq 0, \\ 0, & \text{otherwise}, \end{cases} \quad (2)$$

which is shown in Fig. 1-2. Taylor approximations of degree 50 at different points are displayed in Fig. 1. By contrast, the Chebyshev approximations of function (2) are shown in Fig. 2. □

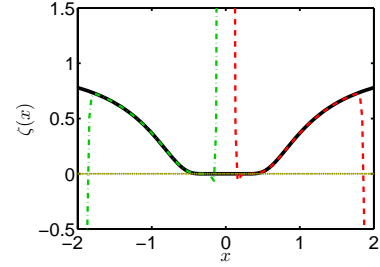

Fig. 1. Example 1: Taylor approximation of degree 50 at different points. The solid black line depicts the original function (2); the dashed red line, the dotted yellow line, and the dotted dashed green line depict the Taylor approximation at $x_0 = 1, 0, -1$, respectively.

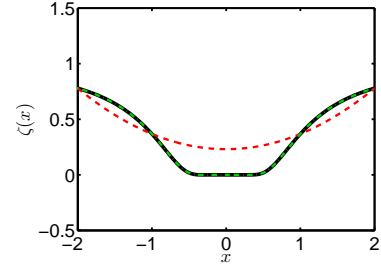

Fig. 2. Example 1: Chebyshev approximations of different degrees. The solid black line depicts the original function (2); the dashed red line and the dotted dashed green line depict the Chebyshev approximation of degree 4 and 50, respectively.

Chebyshev approximation is an interpolation method via Chebyshev polynomials and non-equidistant points [24]. Specifically, let us define the $n$-th order polynomial

$$P_n(x) = \cos(n \arccos(x)), \quad (3)$$

from which one has the following polynomials for $n = 0, 1, 2, 3, \ldots$: $P_0(x) = 1$, $P_1(x) = x$, $P_2(x) = 2x^2 - 1$, $P_3(x) = 4x^3 - 3x, \ldots$. An iterative formula to compute these polynomials is given as:

$$P_{n+1}(x) = 2x P_n(x) - P_{n-1}(x), \text{ with } n \geq 1. \quad (4)$$

Note that each polynomial $P_n$ has exactly $n$ zeros and $n+1$ extrema on the interval [-1,1]. The corresponding values of $x$ at zeros and extrema are given by

$$\begin{aligned} x_z &= \cos\left(\frac{\pi(j - \frac{1}{2})}{n}\right), \quad j \in [1, n], \\ x_e &= \cos\left(\frac{\pi j}{n}\right), \quad j \in [0, n]. \end{aligned} \quad (5)$$

Let $x_{z_n}$ be the $n$-th zero with respect to $x_z$, and define coefficients $c_j$ with $j = 0, 1, \ldots, k-1$ as

$$c_j = \frac{2}{k}\sum_{n=1}^{k} \zeta(x_{z_n})P_j(x_{z_n})$$
$$= \frac{2}{k}\sum_{n=1}^{k} \zeta\left[\cos\left(\frac{\pi(n-\frac{1}{2})}{k}\right)\right]\cos\left(\frac{\pi j(n-\frac{1}{2})}{k}\right), \quad (6)$$

where $\zeta$ is the nonlinear function to be approximated. From [25], the *Chebyshev approximation formula* (without the remainder) for the non-polynomial function $\zeta$ is

$$C^k(\zeta) = \sum_{n=0}^{k-1} c_n P_n(x) - \frac{1}{2}c_0, \quad (7)$$

where $k$ denotes the truncation degree, and $\zeta(x) = C^k(\zeta(x))$ for $x$ equal to any zero of $P_k(x)$. This approximation is applicable to an arbitrary interval $I^* = [a,b]$, which can be achieved by the following transformation:

$$\hat{x} = \frac{x - \frac{1}{2}(b+a)}{\frac{1}{2}(b-a)}. \quad (8)$$

*Remark 3:* It is worth noting that Chebyshev polynomials given by (7) provide an efficient way to approximate non-polynomial functions, and can be very close to the *minimax polynomial*, which is the best polynomial approximation with the smallest maximum deviation. □

For the ease of understanding, an example of Chebyshev approximation is provided for illustration.

*Example 2:* Consider a function $\zeta : \mathbb{R} \to \mathbb{R}^+$ given by

$$\zeta(x) = e^x \quad (9)$$

with selected interval $I = [-1, 1]$ and a truncation degree $k = 4$, we first compute the Chebyshev polynomials according to (4) up to $P_3(x)$, and the coefficients $c_j$ given by (6) as $c_0 \approx 5.0643$, $c_1 \approx 2.2606$, $c_2 \approx 0.5429$ and $c_3 \approx 0.0876$. Then, the Chebyshev approximation is expressed by

$$C^4(e^x) = \sum_{n=0}^{3} c_n P_n(x) - \frac{1}{2}c_0$$
$$\approx 0.9891x^3 + 0.9979x^2 + 0.5539x + 0.1773.$$

We also calculate the Chebyshev approximation with $k = 4$ on $I^* = [-3, 3]$. Via the transformation (8), the approximation results are shown in Fig. 3. □

### C. Problem Formulation

In this paper, we aim to estimate the DA of the origin via polynomial Lyapunov functions. Specifically, let $V(x)$ be a Lyapunov function of system (1) for the origin, which satisfies

$$\forall x \in \mathbb{R}^n_0 : V(x) > 0, \ V(0_n) = 0, \ \lim_{\|x\| \to \infty} V(x) = \infty, \quad (10)$$

and the time derivative of $V(x)$ along the trajectories of (1) is locally negative definite [26]. To this end, we introduce the sublevel set of $V(x)$ as

$$\mathcal{V}(c) = \left\{x \in \mathbb{R}^n : V(x) \leq c\right\}, \quad (11)$$

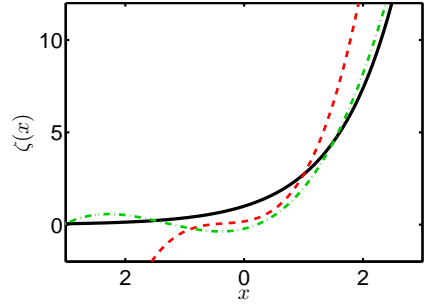

Fig. 3. Example 2: The Chebyshev approximation $C^4(e^x)$. The solid black line depicts the original function (9); the dashed red line and the dotted dashed green line depict the Chebyshev approximation over $[-1, 1]$ and $[-3, 3]$, respectively.

where $c \in \mathbb{R}^+$. For system (1), $\mathcal{V}$ is an estimate of $\mathcal{D}$ if

$$\forall x \in \mathcal{V}(c) \setminus \{0\} : \ \dot{V}(x) < 0. \quad (12)$$

Let us propose the main problems we are concerned with: Find a polynomial Lyapunov function $V(x)$ and a positive scalar $c$ such that the estimate of the DA is maximized under certain selected criteria, i.e., solving

$$\mu = \sup_{c,\ V} \ \rho(\mathcal{V}(c)) \quad (13)$$
$$\text{s.t. (10)} - \text{(12) hold,}$$

where $\rho$ is a measure of $\mathcal{V}(c)$ as a user-defined criteria, e.g., the volume of $\mathcal{V}(c)$.

### III. MAIN RESULTS

In this section, we first approximate the non-polynomial terms by using Chebyshev approximation without the remainder, and a method is proposed via the higher order derivatives of Lyapunov functions. Then, the remainder of Chebyshev approximation is considered. An uncertain polynomial system is reformulated by parameterizing the remainder, and an approach is proposed by estimating the DA of the uncertain polynomial system.

### A. Chebyshev Approximation without the Remainder

Since the DA of non-polynomial system (1) is hard to compute, the basic idea behind approximation methods is to compute the DA of the approximated system

$$\dot{x} = f(x(t)) + \sum_{i=1}^{r} g_i(x(t))C_i^k(\zeta_i(x_{a_i})), \quad (14)$$

whose DA of the origin denotes $\hat{\mathcal{D}}$. For the brevity of notations, let $\hat{h}(x) = f(x(t)) + \sum_{i=1}^{r} g_i(x(t))C_i^k(\zeta_i(x_{a_i}))$. Define the *standard Lyapunov function* as $V(x)$ satisfying (10)-(12). The $m$-th time derivative of $V$ along the trajectories of (14) is denoted as $V^{(m)}(x)$[1].

---

[1] $\dot{V}(x) = V^{(1)}(x) = \langle \nabla V, \hat{h}(x) \rangle$ where $\langle \cdot, \cdot \rangle$ is the inner product of two vectors, $\nabla V$ is the gradient of $V$, i.e., $\nabla V = (\frac{\partial V}{\partial x_1}, \ldots, \frac{\partial V}{\partial x_n})$; $V^{(2)}(x) = \langle \nabla V^{(1)}, \hat{h}(x) \rangle$; $V^{(m)}(x) = \langle \nabla V^{(m-1)}, \hat{h}(x) \rangle$.

*Lemma 1 ([27]):* Suppose a positive definite function $V(x)$ satisfying (10). Then, the origin is globally asymptotically stable if there exist some positive scalars $e_1, e_2, \ldots, e_{m-1}$, such that

$$V^m(x) + e_{m-1}V^{m-1}(x) + \cdots + e_1\dot{V}(x) < 0. \quad (15)$$

The above result generalizes the case of [28] ($m = 3$ in (15)), and allows the existence of non-monotonic Lyapunov functions, i.e., $\dot{V} > 0$ holds for some time intervals. Based on this result, in [18], Ahmadi pointed out that as long as condition (15) holds, there exists a standard Lyapunov function for system (14):

*Lemma 2 ([18]):* If condition (15) is satisfied, then there exist different functions $V_1(x), \ldots, V_m(x)$ such that

$$V_m^{(m-1)}(0) + V_{m-1}^{(m-2)}(0) + \cdots + \dot{V}_2(0) + V_1(0) = 0, \quad (16)$$

and

$$V_m^{(m-1)}(x) + V_{m-1}^{(m-2)}(x) + \cdots + \dot{V}_2(x) + V_1(x) > 0, \quad (17)$$

$$V_m^{(m)}(x) + V_{m-1}^{(m-1)}(x) + \cdots + V_2^{(2)}(x) + \dot{V}_1(x) < 0, \quad (18)$$

for all $x \neq 0$. $\square$

This result is important in the sense that it provides a convex way to construct a standard Lyapunov function

$$W(x) = V_m^{(m-1)}(x) + V_{m-1}^{(m-2)}(x) + \cdots + \dot{V}_2(x) + V_1(x) \quad (19)$$

even if non-monotonic Lyapunov functions exist. In addition, this result is better than the method directly using Lyapunov stability theorem because $W(x)$ involves not only the current state $x$, but also the future value of $\hat{h}(x)$ (For more details, please find in [17], [29]). Let $\Lambda(x) = (V_1(x), V_2(x), \ldots, V_m(x))$, we denote $W(x)$ as $W(\Lambda, \hat{h})$.

In order to use the higher order derivatives of Lyapunov functions to estimate the DA, we would like to introduce a transformed version of Real Positivestellensatz [30], which connects the cone of local SOS with the positivity over a semialgebraic set.

*Lemma 3 ([23]):* For polynomials $a_1, \ldots, a_m, b_1, \ldots, b_l$ and $p$, define a set

$$\begin{aligned} \mathcal{B} = \{x \in \mathbb{R}^n : &a_i(x) = 0, \ \forall i = 1, \ldots, m, \\ &b_i(x) \geq 0, \ \forall j = 1, \ldots, l\}. \end{aligned} \quad (20)$$

Let $\mathcal{B}$ be compact. Condition $\forall x \in \mathcal{B} : p(x) > 0$ can be established if the following condition holds:

$$\begin{cases} \exists r_1, \ldots, r_m \in \mathcal{P}, \ s_1, \ldots, s_l \in \mathcal{P}_0^{\text{SOS}} \\ p - \sum_{i=1}^{m} r_i a_i - \sum_{i=1}^{l} s_i b_i \in \mathcal{P}_0^{\text{SOS}}, \end{cases} \quad (21)$$

where the set of local SOS $\mathcal{P}_0^{\text{SOS}}$ is defined in Section II.

Now, we propose a condition to estimate $\hat{D}$ based on SOS constraints.

*Lemma 4:* For system (14) and a selected truncation degree $k$, assume there exist polynomial functions $\Lambda(x) = (V_1(x), V_2(x), \ldots, V_m(x))$ and $W(\Lambda, \hat{h})$ defined in (19) satisfying

$$\begin{cases} \forall x \in \mathbb{R}_0^n : \ W(\Lambda, \hat{h}) > 0, \ W(\Lambda(0_n), \hat{h}(0_n)) = 0, \\ \lim_{\|x\| \to \infty} W(\Lambda, \hat{h}) = \infty, \end{cases} \quad (22)$$

consider a positive scalar $c \in \mathbb{R}^+$ and polynomial functions $s(x)$ such that

$$\forall x \in \mathbb{R}_0^n : \ \begin{cases} -\psi(x, c, s(x)) \in \mathcal{P}_0^{\text{SOS}} \\ s(x) \in \mathcal{P}_0^{\text{SOS}}, \end{cases} \quad (23)$$

where

$$\psi(x, c, s(x)) = r(x) + s(x)(c - W(\Lambda, \hat{h})), \quad (24)$$

and $r(x) = \langle \nabla W(\Lambda, \hat{h}), \hat{h} \rangle$. Then,

$$\mathcal{W}(c) = \{x | \ W(\Lambda, \hat{h}) \leq c\} \subseteq \hat{\mathcal{D}}. \quad (25)$$

*Proof:* Suppose (23) holds, then $-\psi(x, c, s(x))$ and $s(x)$ are local SOS. From Lemma 3, it yields that there exist polynomial functions $V_1(x), V_2(x), \ldots, V_m(x)$ such that

$$\dot{W}(\Lambda, \hat{h}) < 0, \quad (26)$$

for all $x \in \{x \in \mathbb{R}_0^n : c - W(\Lambda, \hat{h}) \geq 0\} \setminus \{0_n\}$, i.e., (12) holds. Therefore, from Lyapunov stability theorem, $\mathcal{W}(c)$ is an estimate of the DA, which completes this proof. $\square$

From the above result, we know $\mathcal{W}(c)$ is an under-estimate of $\hat{\mathcal{D}}$. But additional questions arise: How can one establish condition (23)? How can one enlarge the $\mathcal{W}(c)$ such that a best approximation can be obtained? To answer these questions, we first define

$$\gamma = \sup c \quad (27)$$

such that (23) holds.

### B. SMR based Quasi-Convexification

Let us observe that the condition (23) is non-convex due to the product of a local SOS $s(x)$ and a scalar $c$. In literature, *Bisection methods* and *iterative SOS programming* are common ways to solve this problem, but these approaches are not applicable with local SOS constraints.

In this subsection, a specific SMR is introduced for the local SOS, i.e., $p_0(x) \in \mathcal{P}_0^{\text{SOS}}$, and we will show that the non-convex problem (23) can be transformed into a generalized eigenvalue problem, which is quasi-convex. Let us first develop the SMR for local SOS: Consider a polynomial $p_0(x)$ of degree $\deg(p_0)$ without the constant and linear terms, we have $p_0(x) \in \mathcal{P}_0^{\text{SOS}}$, whose SMR is given by:

$$p_0(x) = (*)^T (\bar{P}_0 + L(\delta))\phi(n, d_{p_0}), \quad (28)$$

where $(*)^T AB$ is short for $B^T AB$ introduced in Section II; $\bar{P}_0$ is called the SMR matrix of $p_0(x)$; $n$ is the number of variables; $d_{p_0}$ denotes the smallest integer not less than $\frac{\deg(p_0)}{2}$, i.e., $d_{p_0} = \lceil \frac{\deg(p_0)}{2} \rceil$; $\phi(n, d_{p_0}) \in \mathbb{R}^{l(n, d_{p_0})}$ is a power vector involving all monomials whose degree is less or

equal to $d_{p_0}$ without degree 0; and $L(\delta)$ is a parameterization of the affine space

$$\mathscr{L} = \{L(\delta) \in \mathbb{R}^{l(n,d_{p_0}) \times l(n,d_{p_0})} : L(\delta) = L^T(\delta), \\ (*)^T L(\delta) \phi(n, d_{p_0}) = 0\}, \quad (29)$$

in which $\delta \in \mathbb{R}^{\vartheta(n,d_{p_0})}$ is a vector of free parameters. $l(n, d_{p_0})$ and $\vartheta(n, d_{p_0})$ can be calculated similarly to [31] for the case of standard SOS. For the ease of understanding, an illustration of an SMR is given.

*Example 3:* Given the polynomial $p_1(x) = 9x^6 + 8x^5 + 3x^4 + 5x^2$, we have $d_{p_1} = 3$, $n = 1$ and $\phi(n, d_{p_1}) = (x^3, x^2, x)^T$. Then, $p_1(x)$ can be written as follows,

$$P = \begin{pmatrix} 9 & 4 & 0 \\ 4 & 3 & 0 \\ 0 & 0 & 5 \end{pmatrix}, L(\delta) = \begin{pmatrix} 0 & 0 & -\delta \\ 0 & 2\delta & 0 \\ -\delta & 0 & 0 \end{pmatrix}.$$

$\square$

By exploiting the representation introduced in (28), we have the following expressions of SMR:

$$V_i(x) = (*)^T \bar{V}_i \phi(n, d_{v_i}), \ i = 1, \ldots, m \quad (30)$$
$$\hat{h}(x) = (*)^T \bar{H} \phi(n, d_h), \quad (31)$$
$$W(\Lambda, \hat{h}) = (*)^T \overline{W}(\bar{V}_1, \ldots, \bar{V}_m, \bar{H}) \phi(n, d_w), \quad (32)$$
$$s(x) = (*)^T \bar{S} \phi(n, d_a), \quad (33)$$
$$\psi(x, c, s) = (*)^T \bar{\Psi}(\delta, c, \bar{S}) \phi(n, d_\psi), \quad (34)$$

where $\bar{V} \in \mathbb{R}^{l(n,d_{v_i}) \times l(n,d_{v_i})}, \delta \in \mathbb{R}^{\vartheta(n,d_\psi)}$ is a vector of free parameters, and $\overline{W} \in \mathbb{R}^{l(n,d_w) \times l(n,d_w)}, \bar{S} \in \mathbb{R}^{l(n,d_s) \times l(n,d_s)}$ and $\bar{\Psi}(\delta, c, \bar{S}) \in \mathbb{R}^{l(n,d_\psi) \times l(n,d_\psi)}$ are symmetric matrices. Let $\bar{R}(\delta)$, $\Gamma_1(\bar{S})$ and $\Gamma_2(\bar{S})$ be SMR matrices of $r(x)$, $s(x)$ and $W(\Lambda, \hat{h})s(x)$, respectively, with respect to the power vector $\phi(n, d_\psi)$. From (24), it yields

$$\bar{\Psi}(\delta, c, \bar{S}) = \bar{R}(\delta) + c\Gamma_1(\bar{S}) - \Gamma_2(\bar{S}),$$

where $\delta \in \mathbb{R}^{\vartheta(n,d_\psi)}$ is a vector of free parameters. The following result finds the largest estimated DA by solving a generalized eigenvalue problem (GEVP).

*Theorem 1:* Consider positive scalars $\eta_1$, $\eta_2$, a truncation degree $k$, and polynomials $V_1, \ldots, V_m$ fulfilling (22), the lower bound of $\gamma$ can be calculated by

$$\tilde{\gamma} = -\frac{\tilde{e}}{\eta_1 + \eta_2 \tilde{e}} \quad (35)$$

where $\tilde{e}$ is the solution of the GEVP

$$\tilde{e} = \inf_{\delta, \ e, \ \bar{S}} e \\ \text{s.t.} \begin{cases} \eta_1 + \eta_2 e > 0, \\ \bar{S} > 0, \\ e\Gamma(\bar{S}) > \bar{R}(\delta) - \Gamma_2(\bar{S}), \end{cases} \quad (36)$$

where $\Gamma(\bar{S})$ is the SMR matrix of $\eta_1 s(x) + \eta_2 W(\Lambda, \hat{h})s(x)$ with respect to the power vector $\phi(n, d_\psi)$.

*Proof*: In this proof, we first demonstrate that 1) (36) is a GEVP. Secondly, we show that 2) $\tilde{\gamma}$ in (35) is the lower bound of $\gamma$.

1) Optimization (36) is a GEVP: From (32), $\overline{W}$ is the SMR matrix of $W(\Lambda, h)$, and we also have that $\Gamma_1(\bar{S})$ and $\Gamma_2(\bar{S})$ are SMR matrices of $s(x)$ and $W(\Lambda, \hat{h})s(x)$, respectively, with respect to the power vector $\phi(n, d_\psi)$. From [32], one has $\Gamma > 0$ if $\overline{W} > 0$ and $\bar{S} > 0$. Therefore, (36) is a GEVP.

2) $\tilde{\gamma}$ in (35) is the lower bound of $\gamma$: According to the last inequality of (36), we have

$$\widetilde{\Psi}(\delta, c, \bar{S}) = \bar{R}(\delta) - e\Gamma(\bar{S}) - \Gamma_2(\bar{S}) \\ < 0.$$

Considering (34) and

$$\tilde{\psi}(x, c, s(x)) = r(x) - W(\Lambda, \hat{h})s(x) \\ - e(\eta_1 + \eta_2 W(\Lambda, \hat{h}))s(x),$$

we rewrite $\tilde{\psi}(x, c, s(x))$ as the following form:

$$\tilde{\psi}(x, c, s(x)) = \tilde{\psi}(x, \frac{-e}{\eta_1 + \eta_2 e}, (\eta_1 + \eta_2 e)s(x)).$$

Let us observe that the function $-e/(\eta_1 + \eta_2 e)$ is a monotonically decreasing function, and this function maps from the set $(-(\eta_1/\eta_2), 0]$ into the set $[0, +\infty)$. Thus, (35) is the lower bound of $\gamma$. $\square$

For more details of GEVP, please see the book [32]. The above theorem provides a method to approximate $\hat{D}$ with fixed $V_1, \ldots, V_{m-1}$ and thus fixed $W$. To find optimal $V_1, \ldots, V_{m-1}$ for further enlarging $\mathcal{W}(c)$, the following method is proposed:

*Proposition 1:* Assume that there exist a polynomial $s \in \mathcal{P}_0^{\text{SOS}}$ and a positive scalar $c$, such that

$$\epsilon = \inf_{\bar{V}_1, \ldots, \bar{V}_m} \text{trace}(\overline{W}) \\ \text{s.t.} \begin{cases} W(\bar{V}_1, \ldots, \bar{V}_m, x) \in \mathcal{P}_0^{\text{SOS}} \\ -\psi(x, \bar{V}_1, \ldots, \bar{V}_m, c, s(x)) \in \mathcal{P}_0^{\text{SOS}}, \end{cases} \quad (37)$$

with fixed $c$ and $s(x)$, where $\overline{W}(\bar{V}_1, \ldots, \bar{V}_m, \bar{H})$ and $\bar{H}$ are introduced in (32) and (31). Then, $\mu = \frac{\gamma}{\epsilon}$ is an underestimate of $\text{vol}(\hat{D})$.

*Proof*: Similar arguments can be given by using SMR and Real Positivestellensatz as in the proofs of Lemma 4 and Theorem 1. We omit the proof due to limited space. $\square$

*Remark 4:* It is worth noting that to solve Problem 1, one can use an iterative algorithm, which includes step 1: Use Theorem 1 to enlarge the estimate $\mathcal{W}(c)$ with selected $V_1, \ldots, V_m$; and step 2: Use Proposition 1 to search for optimal $V_1, \ldots, V_m$, thus an optimal $W(\Lambda, \hat{h})$ such that the estimate $\mathcal{W}(c)$ can be further enlarged. $\square$

### C. Chebyshev Approximation with the Remainder

In this subsection, we consider the remainder of Chebyshev approximation, and based on the parameterization of this remainder, a method is proposed for directly estimating $\mathcal{D}$ rather than $\hat{\mathcal{D}}$. Specifically, the Chebyshev approximation of non-polynomial function $\zeta_i$ with the remainder on the interval [-1,1] is

$$\zeta_i(x_{a_i}) = C^k\left(\zeta_i(x_{a_i})\right) + \xi_i \frac{\prod_{i=0}^{k-1}(x - x_{z_i})}{k!}, \quad (38)$$

where $C^k$ is given in (7) and $x_{z_i}$ is the $i$-th zero given in (5). For the approximation on an arbitrary closed interval $[a, b]$, let us recall (8) in Section II. Without loss of generality, we consider the approximation on $[-1, 1]$ in this subsection.

*Remark 5:* The Chebyshev remainder, $\zeta_i(x_{a_i}) - C^k(\zeta_i(x_{a_i}))$ is parameterized by $\xi_i$, which is in the hyper-rectangle

$$\Xi = [\underline{\tau}_1, \overline{\tau}_1] \times \cdots \times [\underline{\tau}_r, \overline{\tau}_r] \quad (39)$$

and $\underline{\tau}_i, \overline{\tau}_i \in \mathbb{R}$, $i = 1, \ldots, r$, are chosen as

$$\underline{\tau}_i \leq \left. \frac{d^k \zeta_i(x_{a_i})}{dx_{a_i}^k} \right|_{x_{a_i} = \iota} \leq \overline{\tau}_i, \quad (40)$$

for all $\iota \in \mathcal{I}$, where $\mathcal{I}$ is the set of interest, which is selected in a sublevel set $\mathcal{V}_{x_{a_i}}$ that we will introduce later. Note that the remainder expressed in (38) is called the Lagrange form. For other mean-value forms, e.g., the Cauchy form, it also has the property (40) [25]. □

For the brevity of notations, let us denote $\xi = (\xi_1, \ldots, \xi_r)^T$, and for a chosen Lyapunov function $V(x)$, we introduce the following polynomials:

$$u(x) = \langle \nabla V(x), f(x) + \sum_{i=1}^{r} g_i(x) C_i^k(\zeta_i(x_{a_i})) \rangle, \quad (41)$$

$$q_i(x) = \langle \nabla V(x), g_i(x) \frac{\prod_{j=0}^{k-1}(x - x_{z_j})}{k!} \rangle, \quad (42)$$

$$q(x) = (q_1(x), \ldots, q_r(x))^T. \quad (43)$$

Based on Lemma 3, an approach is provided to compute the estimate of $\mathcal{D}$.

*Theorem 2:* Given a truncation degree $k$, and provided that there exist a polynomial functioin $V(x)$ satisfying (10), a polynomial $d(x) \in \mathcal{P}_0^{\text{SOS}}$, and $\hat{c}$ is the optimum of the following polynomial optimization

$$\hat{c} = \sup_{c, d} c$$
$$\text{s.t.} \begin{cases} -\sigma(x, c, d(x)) \in \mathcal{P}_0^{\text{SOS}} \\ \forall x \in \mathcal{V}(c) \setminus \{0\} \\ \forall \xi_i \in \text{ver}(\Xi), \ i = 1, \ldots, r, \end{cases} \quad (44)$$

where

$$\sigma(x, c, d(x)) = u(x) + q(x)^T \xi + d(x)(c - V(x)), \quad (45)$$

functions $u$ and $q$ are introduced by (41)-(43), a local SOS polynomial $d(x) \in \mathcal{P}_0^{\text{SOS}}$ and $\text{ver}(\Xi)$ is the set of vertices of $\Xi$. Then, $\mathcal{V}(\hat{c}) \subseteq \mathcal{D}$.

*Proof:* Assume that (44) holds, we have that $-\sigma(x, c, d(x))$ and $d(x)$ are local SOS. From Lemma 3, it yields that

$$\forall \xi_i \in \text{ver}(\Xi) : \ u(x) + q(x)^T \xi < 0, \quad (46)$$

for all $x$ in $\{x \in \mathbb{R}^n : c - V(x) \geq 0\} \setminus \{0\}$. From (38) and (40), we obtain that for all $x \in \mathcal{V}(c)$, there exists $\bar{\xi}_i \in [\underline{\tau}_i, \overline{\tau}_i]$ such that $\zeta_i$ in (38) can be expressed as

$$\zeta_i(x_{a_i}) = C^k\left(\zeta_i(x_{a_i})\right) + \bar{\xi}_i \frac{\prod_{i=0}^{k-1}(x - x_{z_i})}{k!}. \quad (47)$$

Thus, from (41)-(43) it results in: For all $x \in \mathcal{V}(c)$, there exist $\bar{\xi}_i \in \Xi$ for all $i = 1, \ldots, r$ and $\bar{\xi} = (\bar{\xi}_1, \ldots, \bar{\xi}_r)^T$ such that

$$\dot{V}(x, \bar{\xi}) = r(x) + q(x)^T \bar{\xi}. \quad (48)$$

Since $\dot{V}(x, \bar{\xi})$ is affine in $\bar{\xi}$ while $\Xi$ is a sort of convex polyhedron, from (46)-(48), we have that

$$r(x) + q(x)^T \bar{\xi} < 0, \quad (49)$$

for all $\bar{\xi}_i \in \Xi$ and for all $x$ in $\{x \in \mathbb{R}^n : c - V(x) \geq 0\} \setminus \{0\}$, i.e., (12) holds. Therefore, it finally yields that $\mathcal{V}(\hat{c})$ is an estimate of the DA with the truncation degree $k$, which ends this proof. □

*Remark 6:* For this result, it is worth noting that

- Theorem 2 provides a method to compute the DA of non-polynomial systems via Chebyshev approximation with the remainder. A standard Lyapunov function $V(x)$ is used rather than $W(x)$ given in (19), because $\dot{W}$ is not affine in $\xi$ when higher order derivatives of Lyapunov functions are considered. In other words, higher order derivatives of Lyapunov functions are not applicable in this case.
- Condition (44) can also be transformed into a GEVP by using SMR technique as Theorem 1. By fixing $c$ and $d(x)$, and searching an optimal $V(x)$ as Proposition 1, one can further reduce the conservativeness and get a larger estimate of the DA. □

## IV. EXAMPLES

The computation is executed using MATLAB 2016a on a standard laptop with an 8GB DDR3 RAM and an Intel Core i7-4712MQ processor. We use MATLAB toolboxes SeDuMi, SMRSOFT and SOSTOOLS for solving LMIs and SOS programmings.

### A. Example 4

Consider the non-polynomial function $\zeta : \mathbb{R} \to \mathbb{R}^+$ as

$$\zeta(x) = \sqrt{|e^x \cos(x)|}, \quad (50)$$

which is continuous on $\mathbb{R}$, and is not differentiable at the points $x_{u_n} = n\pi/2$, $n \in \mathbb{Z}$. Thus, we select $x_0 = 0$ and $I = [-1, 1]$. From Fig. 4-5, it is obvious that Chebyshev approximation is better than Taylor approximation both in accuracy and efficiency.

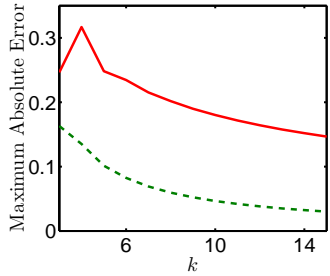

Fig. 4. Example 4: The maximum absolute error corresponding to the truncation degree $k$. The solid red line depicts Taylor approximation; the dashed green line depicts Chebyshev approximation.

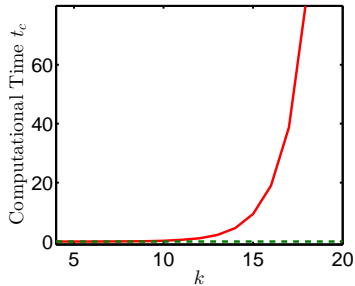

Fig. 5. Example 4: The computational time $t_c$ [sec] corresponding to the truncation degree $k$. The solid red line depicts Taylor approximation; the dashed green line depicts Chebyshev approximation.

### B. Example 5: A Case without the Remainder

We extend Example 3 and consider the following system:

$$\begin{aligned}\dot{x}_1 &= -x_1 + x_1^2 - x_1 x_2 - x_1^3 \\ \dot{x}_2 &= -x_2 + x_1 x_2 + \sqrt{|e^{x_1}\cos(x_1)|} - 1.\end{aligned} \quad (51)$$

We would like to check whether the advantages of Chebyshev approximation can be found in the estimate of the DA of non-polynomial systems. First, we select $m = 2$, the truncation degree $k = 4$, $W(x) = \dot{V}_2(x) + V_1(x)$ and use Theorem 1 and Proposition 1 to find a $V_1(x)$ of degree 1 and $V_2(x)$ of degree 2. The computational result is shown in Fig. 6 and we compare it to the method using standard Lyapunov functions of degree 4. It is not hard to find that the estimate via Taylor approximation is out of the boundary of the exact DA, due to a comparatively large error of Taylor approximation. In addition, we show the efficiency of the proposed algorithm given in Remark 4 by Table I: The proposed method is much faster due to a smaller number of decision variables. Therefore, this example also demonstrates that the proposed method is advanced both in accuracy and efficiency compared to Taylor approximation.

### C. Example 6: A Case with the Remainder

Implementations can be extended to a 3D example:

$$\begin{aligned}\dot{x}_1 &= x_2 + x_3^2 \\ \dot{x}_2 &= x_3 - x_1^2 - x_1\sin(x_1) \\ \dot{x}_3 &= -x_1 - 2x_2 - x_3 + x_2^3 + \ln\left(\tfrac{1+x_3}{1-x_3}\right)^{\tfrac{1}{10}},\end{aligned}$$

TABLE I
THE COMPUTATIONAL TIME $t_c$ [sec] FOR THE DIFFERENT METHODS, THE NUMBER OF ITERATION $n_t$, AND THE TRUNCATION DEGREE $k$.

| | $k=4$ | | | $k=6$ | | |
|---|---|---|---|---|---|---|
| | $n_t$=5 | $n_t$=10 | $n_t$=20 | $n_t$=5 | $n_t$=10 | $n_t$=20 |
| This method | 15.12 | 28.51 | 58.17 | 22.32 | 46.91 | 87.47 |
| [23] | 38.63 | 64.16 | 174.43 | 68.81 | 150.53 | 332.31 |

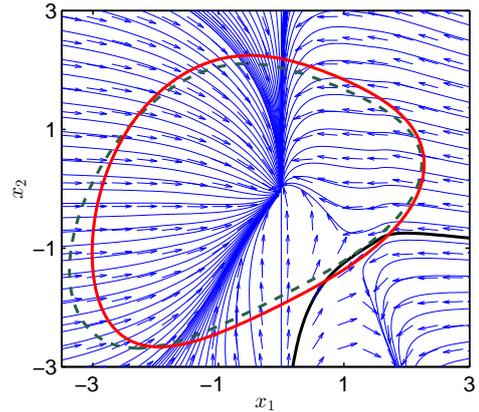

Fig. 6. Example 5: The computational results with $k = 4$ and $m = 2$. The solid black line depicts the boundary of the exact DA. Some trajectories are shown as blue lines. The dashed green line indicates the estimate by using the higher-order derivatives of Lyapunov function and Chebyshev approximation. The solid red line depicts the estimate by using a standard Lyapunov function and Taylor approximation as in [23].

with $g_1 = x_1$, $g_2 = 0.1$, $\zeta_1 = \sin x_1$, $\zeta_2 = \ln\left(\tfrac{1+x_3}{1-x_3}\right)$. In this case, we use a standard Lyapunov function with the truncation degree $k = 3$ and consider the remainder of approximations. According to Theorem 2, we use a standard Lyapunov function of degree 4 to estimate the DA, and compare it with the method proposed in [23]. Note that both methods are inner-approximation methods, as shown in Fig. 7, the proposed method has a better performance due to a more accurate estimate via Chebyshev approximation.

## V. CONCLUSION AND DISCUSSION

An approach is proposed for estimating the DA of a class of non-polynomial systems by using Chebyshev approximation. It is shown that the proposed approach is better both in accuracy and efficiency compared to Taylor approximation. Firstly, for Chebyshev approximation without the remainder, a condition is provided for estimating the DA via higher-order derivatives of Lyapunov functions. In addition, a problem of quasi-convex optimization is constructed based on the square matrix representation. Finally, for Chebyshev approximation with the remainder, an uncertain polynomial system is built which is linearly influenced by parameters constrained in a hyper-rectangle. Then, the lower bound of the largest estimate of the DA can be obtained by using a standard Lyapunov function.

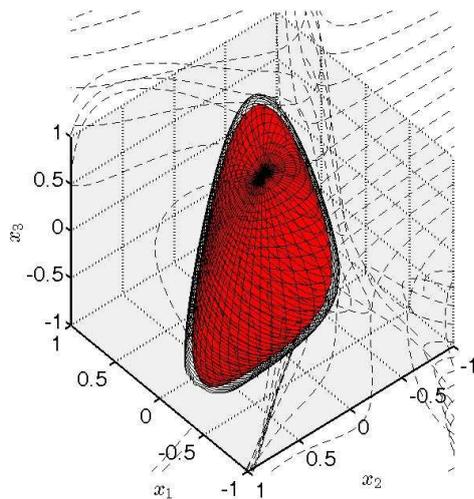

Fig. 7. Example 6: The computational results with $k = 3$. The red region indicates the estimate of DA using Taylor approximation with $\deg(V_1) = 4$. The solid black lines indicate the estimate of DA using Chebyshev approximation with $\deg(V_2) = 4$. Some dashed lines depicting $\dot{V}_2(x) = 0$ are also shown.

The conservativeness of this method originates from the fact that (36) only provides a suboptimal solution. To cope with this issue, a convex method is proposed based on the moment theory [11], [12], to which our future efforts will be devoted. To further reduce the conservativeness, we will focus on using rational Lyapunov functions and multi-sublevel sets methods [9], [33].